\documentclass[epj]{svjour}
\usepackage{graphics}
\begin{document}
\title{Molecular dynamics simulation for the baryon-quark phase transition
at finite temperature and density}
\author{Yuka Akimura\inst{1,} \inst{2} \and Toshiki Maruyama\inst{2}
 \and Naotaka Yoshinaga\inst{1} \and Satoshi Chiba\inst{2}  
}                     
\institute{Department of physics, Saitama University, 255 Shimo-Okubo, Sakura-Ku,
 Saitama City, Saitama 338-8570, Japan,
\email{akimura@tiger04.tokai.jaeri.go.jp}
 \and Advanced Science Research Center, Japan Atomic Energy Research Institute,
 Tokai, Ibaraki 319-1195, Japan
 }
\date{Received: date / Revised version: date}
\abstract{We study the baryon-quark phase transition
in a molecular dynamics (MD) of quark degrees of freedom
at finite temperature and density.
The baryon state at low density and temperature, and the deconfined quark state
at high density and temperature are reproduced.
We investigate the equations of state of
matters with different $u$-$d$-$s$ compositions.
Then we draw phase diagrams in the temperature-density plane by this simulation.
It is found that the baryon-quark transition is sensitive to the quark width.
\PACS{
      {PACS-key}{discribing text of that key}   \and
      {PACS-key}{discribing text of that key}
     } 
} 
\maketitle
\section{Introduction}\label{intro}
It is expected that quarks are observed as deconfined states
at extreme environment:
at high density and/or high temperature, where baryons or thermal pions overlap
with each other, disappearance of hadron boundaries may give rise to the
quark gluon plasma (QGP).
Also the change of coupling constants and the quark mass
due to the nature of the quantum chromodynamics (QCD),
is thought to be one of the origins of the QGP \cite{Coln}.
Recently many theoretical calculations are attempted to draw a QCD phase diagram
\cite{Sch99,Kogt04,Taw04}.
The critical temperature of the hadron-quark phase transition is predicted
to be around 150 MeV by lattice simulations \cite{Kogt83,Step04}.
For a finite density system, the bag model predicts in a qualitative
study \cite{Satz98}
that the critical density lies at a few times the normal nuclear saturation density.
Many efforts are devoted to experimental search for the QGP
by using high energy accelerators.
There, many indirect signals of the QGP have been observed \cite{JPsi,Azi,Qcond}
but a definite conclusion has not been obtained
since theoretical studies are not
enough to characterize the properties of the QGP and hadron gas at present.
It is not clear how hadron matter changes to quark matter
and how the physical property changes at the hadron-quark transition.

The Lattice simulation based on the first principle
is the most reliable method for QCD.
It has been applied to high-temperature and low-density systems
under some approximations.
At high density, however, the complex fermion determinant,
or the sign problem, makes the simulation practically impossible \cite{Ejiri04}.
One of the approaches for treating finite density systems is
a mean field theory
which makes the most of its ability for uniform systems.
There are many works with the mean field treatment,
which are incorporated with
the chiral restoration by the Nambu-Jona-Lasinio model \cite{NJL},
the bag model picture \cite{Bag}, the soliton model picture \cite{soltn}, and so on.
The dynamics of hadron-quark transition, however, are not considered in these cases.
In these circumstances, other approaches which are feasible
for treating the finite density
and the dynamics are being awaited.

The molecular dynamics (MD) has been successful
in treating many-body nucleon systems.
To describe the structure and the dynamics of many-body nucleon systems,
several MD models were proposed \cite{Aich,Feld90,Ono92,Maru96}
and obtained some remarkable results.
Advantages of the MD simulation are that one needs very few assumptions a priori
and that a single model can be applied to various problems.
To investigate properties of hadron and quark matter and the dynamics of
transition between these two phases in a unified manner,
MD is a natural framework to be attempted.
In this paper we apply a model similar to quantum molecular dynamics (QMD)
\cite{Aich}
to many-body quark systems,
within a framework of the non-relativistic quark model.
Some pioneering works of applying MD
to quark systems were presented
where transitions caused by increasing density and temperature
were studied
in the view point of many-body dynamics:
in \cite{Bona99,Bona00} the Vlasov equation and the Vlasov+MD approach were used to ge
t the equation of state
of matter.
In \cite{Hof00}
quark MD was applied to heavy-ion collisions
where the creation of particle-antiparticle pairs was taken into account.
In \cite{Maru00}, the color gauge symmetry was treated exactly
and the meson exchange potentials were introduced.
In the above studies  quarks were treated as classical particles and
there was ambiguity in the definition of the ground state of the system.
In the present paper, we propose a model
which has less ambiguity to describe the ground state and to define the temperature.
With this model we study the mechanism of the baryon-quark transition
and draw the phase diagram together with the equation of state (EOS)
for a wide range of baryon density and temperature.

This paper is organized as follows.
In section \ref{basic} we explain basics of the model in detail.
Simulation results for the three kinds of matters:
$ud$ (matter containing the same number of $u$ and $d$ quarks;
 corresponds to symmetric nuclear matter),
$udd$ ($d$ quarks twice the number of $u$ quarks; neutron matter) and
$uds$ (the same number of $u$, $d$ and $s$ quarks; $\Lambda$ matter)
by using two kinds of quark width
 are given in section \ref{result}.
In section \ref{tempera} we apply the model to finite temperature systems
and present phase diagrams.
Summary is given in section \ref{summ}.
\section{molecular dynamics for quark matter}\label{basic}
\subsection{Wave functions and cooling equations}
We start with the total wave function by a direct product of $n=3A$
single particle quark Gaussian wave packets
in coordinate and momentum spaces and
state vectors $\chi$  with a fixed flavor, a color and a spin orientation,
\begin{equation}
\Psi 
=\prod_{i=1}^{n} \frac{1}{(\pi L^2)^{3/4}}
\exp \left[ -\frac{({\bf r}_i-{\bf R}_i)^2}{2L^2}
+\frac{i}{\hbar}{\bf P}_i{\bf r}_i
\right]\chi_i,
\end{equation}
where $n=3A$ ($A$ is the baryon number),
$L$ the fixed width of wave packets,
and ${\bf R}_i$ and ${\bf P}_i$
are the center of the wave packet
of $i$-th quark in coordinate and momentum spaces, respectively.
Instead of antisymmetrization of the total wave function,
the fermionic nature of the system is phenomenologically
treated by introducing the Pauli potential
which acts repulsively between quarks having the same flavor,
 color and spin orientation \cite{Maru97}.
The equations of motion for ${\bf R}_i$ and ${\bf P}_i$ are given by the Newtonian equations,
\begin{equation}
\dot{{\bf R}_i}=\frac{\partial H}{\partial {\bf P}_i}
,\;\;\;
\dot{\bf P}_i=-\frac{\partial H}{\partial {\bf R}_i}.
\label{0cool}
\end{equation}
When we search for the ground state (energy minimum configuration)
of the system, we solve the equations of motion with friction terms,
which we call ``cooling equations of motion'',
\begin{equation}
\dot{{\bf R}_i}=\frac{\partial H}{\partial {\bf P}_i}
+\mu_r\frac{\partial H}{\partial {\bf R}_i},\;\;\;
\dot{\bf P}_i=-\frac{\partial H}{\partial {\bf R}_i}
+\mu_p\frac{\partial H}{\partial {\bf P}_i},
\label{cool}
\end{equation}
where $\mu_r$ and $\mu_p$ are negative frictional coefficients.
The cooling is performed until the particles stop ($\dot{\bf R}_i=0$).
In the ground state, the momenta ${\bf P}_i$ have finite values
because of the momentum dependence of the Pauli potential.

In order to simulate the infinite system (matter)
by using a finite number of quarks,
we use a cubic cell with 26 mirroring cells
under a periodic boundary condition. 
The cell size is chosen to be 6 fm throughout this paper.
\subsection{Effective Hamiltonian}
The effective Hamiltonian consists of three parts as,
\begin{equation}
H=H_0+V_{\rm Pauli}-T_{\rm spur},
\end{equation}
where $H_0$ is the original Hamiltonian expressed as
\begin{equation}
H_0\equiv
\left<\Psi\left|
\sum_{i=1}^{n}
\hat{T}_i
+  \hat  V_{\rm color}+\hat V_{\rm meson}
\right| \Psi \right> ,
\end{equation}
\begin{equation}
E_i\equiv
\left<\Psi\left| \hat{T}_i \right| \Psi \right>
= \frac{{\bf P}_i^2}{2m_i} + \frac{3\hbar^2}{4m_i L^2} + m_i
\label{kinetic}.
\end{equation}
We employ the quark-quark interactions as follows,
\begin{equation}
\hat V_{\rm color}= \frac{1}{2} \sum_{i=1, j \neq i}^{n}
\left( -\sum_{a=1}^8\frac{\lambda_i^a \lambda_j^a}{4}
\left(K\hat{r}_{ij}-\frac{\alpha_s}{\hat{r}_{ij}}
\right) \right),
\end{equation}
\begin{eqnarray}
\hat V_{\rm meson}&=&
  \sum_{{i=1,i \tiny{\in}{l}}}^{n} \left[ \frac{1}{2-\varepsilon}
\left( -\frac{g_{\sigma q}^2}{4\pi} \right)
\left(   \sum_{j \neq {i,j \scriptsize{\in} l}}^{n}
\frac{e^{-\mu_{\sigma}\hat{r}_{ij}}}{\hat{r}_{ij}}\right)^{1-\varepsilon}
\right. \nonumber \\
&+& \left.\frac{1}{2}  \sum_{j \neq i, j\in l}^{n} \left(
\frac{g_{\omega q}^2}{4 \pi}
\frac{e^ {- \mu _{ \omega } \hat{r}_{ij}}}{\hat{r}_{ij}}
+\frac{\sigma_i^3 \sigma_j^3}{4}\frac{g_{\rho q}^2}{4 \pi}
\frac{e^ {- \mu _{ \rho } \hat{r}_{ij}}}{\hat{r}_{ij}} \right) \right]
\nonumber \\
&+& \frac{1}{2} \sum_{i=1, i\in s}^{n} \sum_{j \neq i, j\in s}^{n}
 \frac{g_{\phi q}^2}{4 \pi}
\frac{e^ {- \mu _{ \phi } \hat{r}_{ij}}}{\hat{r}_{ij}},
\label{Vmeson}
\end{eqnarray}
where $\hat{r}_{ij}\equiv |{\bf r}_i-{\bf r}_j|$ and
$l$ means a light flavor $u$ or $d$.
The color dependent interaction $\hat{V}_{\rm color}$ consists of the linear confining
 potential
with an infrared cut-off at 3 fm
and the one gluon exchange potential \cite{Maru00}
with Gell-Mann matrices $\lambda^a$.
To include the antisymmetric effect for the matrix elements of
color space, we use effective values of
$\left<\chi _i| \lambda_i^a \lambda_j^a | \chi _j \right>^{\rm eff}=
4\left<\chi _i| \lambda_i^a \lambda_j^a | \chi _j \right>$
\cite{Maru00}.
The color force is approximately canceled between a colored quark
and a white baryon made of quarks with three colors locating near to each other.
$\hat{V}_{\rm meson}$ consists of  $\sigma$, $\omega$ and
$\rho$ meson exchange potentials which
act among light flavor quarks ($u$ quarks and $d$ quarks),
and the $\phi$ meson exchange potential which
acts among $s$ quarks, where
$\sigma^3$ is the third component of Pauli matrices.
We modify the $\sigma$ exchange potential
of Yukawa-type with a small non-linearity parameter $\varepsilon$.
This corresponds to a density dependent potential or a method
in the relativistic mean field theory (RMF)
where higher order terms in $\sigma$ field are introduced to
reproduce the saturation property for the symmetric nuclear matter.
For simplicity, we have written Eq.\ (\ref{Vmeson}) in the form of an operator.
In practice, however, the power by $1-\varepsilon$ of the $\sigma$ exchange potential
is performed after evaluating the expectation value.
The parameters in the meson exchange potentials are adjusted
to reproduce the baryon-baryon interactions
as described in Sec.\ \ref{basic} C.

Lack of the antisymmetrization is compensated by using the Pauli potential,
\begin{equation}
V_{\rm Pauli}=
\frac{C_{p}}{\left(q_0 p_0 \right)^3}
\exp \left[-\frac{({\bf R}_i-{\bf R}_j)^2}{2q_0^2}
-\frac{({\bf P}_i-{\bf P}_j)^2}{2p_0^2}
\right]\delta_{\chi_i \chi_j} ,
\end{equation}
where $q_0$, $p_0$ and $C_p$ are parameters determined in Sec.\ \ref{basic} C.
Antisymmetrized wave functions are not used
because it takes CPU time proportional to the forth power of a
particle number \cite{Feld90,Ono92}.
On the other hand, MD without antisymmetrization needs CPU times
proportional to the second power of the particle number.
At present, simulation with antisymmetrization is not
practically possible for the system with several hundreds of particles.
Furthermore, the way to antisymmetrize wave functions with a
periodic boundary condition has not been established yet.
We have to subtract the spurious zero point
energy of the center-of-mass motion of clusters
\cite{Ono92,Maru96},
\begin{eqnarray}
T_{\rm spur}&=&\sum_{i=1}^{n} \frac{3\hbar^2}{4m_{i}L^2}
\frac{1}{M_i}, \\
M_i&\equiv& \sum_{j=1}^{n} f_{ij},\\
f_{ij}&\equiv&
\left \{
\begin{array}{ll}
1 & (R_{ij} \le a_f) \\
\exp \left[ - \frac{({R}_{ij}- a_f)^2}{w_f^2} \right]
& (R_{ij} > a_f)
\end{array} \right. ,
\end{eqnarray}
where $R_{ij}\equiv |{\bf R}_i-{\bf R}_j|$ and
$M_i$ is the ``mass number'' of fragment to which the wave packet $i$ belongs,
which is the sum of ``friendship'' $f_{ij}$ with other particles.
To apply the model to various densities,
we introduce a density-normalized parameters of cluster separation
in the friendship as
\begin{equation}
a_{f}=a_{0} \left( \frac{\rho_0}{\rho} \right)^{1/3} ,\ \ \
w_{f}=w_{0} \left( \frac{\rho_0}{\rho} \right)^{1/3} ,
\label{separation}
\end{equation}
where $\rho_0$ means the normal nuclear density $0.17$ fm$^{-3}$.

\subsection{Choice of parameters}
\subsubsection{Quark model parameter and friendship}
We use the constituent quark mass $m_{u,d}$ = 300 MeV for light quarks and
$m_s$ = 500 MeV for $s$ quarks throughout this simulation.
For the color potential, 900 MeV/fm is used for the string tension $K$ and
1.25 for the QCD fine structure constant $\alpha_s$,
 which are typical in quark models \cite{TYos00}.

The parameters in the friendship are chosen as $a_0$ = 0.3 fm and $w_0$ = 0.5 fm,
 so that the sum of the friendship $M_{i}$ $\approx$ 3 when the system clusterizes as
baryons and $M_{i}$ $\approx$ 1 when quarks do not make clusters.

The width of quarks $L$ is the most important parameter in this model
 since it is directly related to the density at which baryon-quark transition occurs.
 We use two different widths by considering the masses of isolated nucleon
$N$ and lambda particle $\Lambda$.
To minimize the energy (mass) of a nucleon, $L$ becomes 0.43 fm.
%
However, the nucleon mass is too large (about 2400 MeV) with this value of $L$.
  The reason of this overestimation is
that the ground-state kinetic energy per quark in a nucleon,
 $\displaystyle\frac{2}{3}\frac{3 \hbar ^2}{4 m_{u,d} L^2}$,
 has a large value of 351 MeV,
while the kinetic energy of a quark in a nucleon
is roughly estimated to be 50 -- 80 MeV from the uncertainty principle.
Therefore, we employ ``effective" widths $L^{\rm eff}$
 in evaluating $E_i$ (Eq.\ (6)) and $T_{\rm spur}$ (Eq.\ (10)).
The ground-state kinetic energy per quark in a nucleon becomes 65 MeV
if we use $L^{\rm eff}=1.0$ fm.

Our first choice is to use $L$ = 0.43 fm, $L^{\rm eff}_{u,d}$ = 1.0 fm
and $L^{\rm eff}_{s}$ = 0.8 fm.
We call this choice as ``width parameter (I)".
This combination
 still gives a slightly large value of the nucleon mass,
 $M_{N}$ $\approx$ 1500 MeV,
 but gives a proper difference of masses
 between $N$ and $\Lambda$.

Once the effective width $L^{\rm eff}$ for kinetic energy is employed,
the nucleon mass does not have a minimum regarding $L$.
Then the second choice is made to give a proper nucleon mass value,
 $M_{N}$ $\approx$ 940 MeV by changing $L$ to be 0.33 fm
(but with the same values of $L^{\rm eff}$ as before).
In this case, the difference of masses between $N$ and $\Lambda$ cannot be set to
 a desired value.
 We call this choice as ``width parameter (II)".
\subsubsection{Pauli potential}
\begin{figure}
\resizebox{0.47\textwidth}{!}{%
\includegraphics{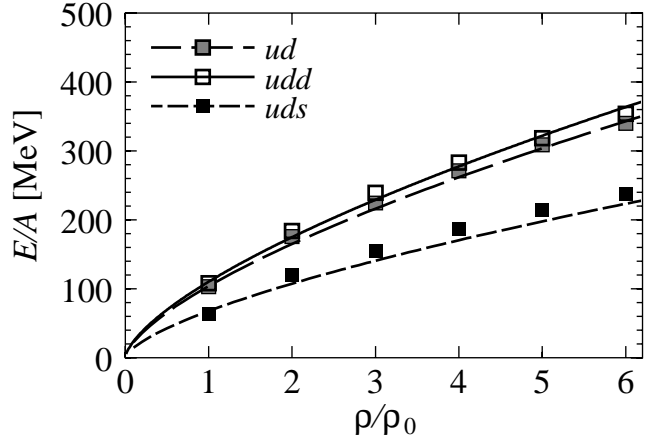}}
\caption{Energy per baryon of free Fermi gas. The lines show
the theoretical exact values of $ud$, $udd$ and
$uds$ matter. The marks are the kinetic energies
calculated by MD with only the Pauli potential.
}
\label{fig:1}       
\end{figure}
For the Pauli potential, the parameters ${C_p,q_0,p_0}$ are determined
by fitting the kinetic energy to the exact Fermi gas at zero temperature.
We determine these values by solving the cooling equation (\ref{cool})
for $ud$ matter, $udd$ matter and $uds$ matter
where only the Pauli potential is considered
\cite{Maru97}.
Figure 1 shows the classical and non-relativistic kinetic energy of the Fermi gas.
The lines indicate the values of the exact Fermi gas energy for $ud$ matter,
$udd$ matter and $uds$ matter.
The squares show the kinetic energy of each matter,
whose values are obtained by using following parameters:
we adopt
$q_0=1.6$ fm, $p_0=120$ MeV and $C_p=131$ MeV for light quarks
and $C_p=79$ MeV for $s$ quarks.
Here three different colors and two different spins are assumed for each matter.
It is seen that the difference between $ud$ matter and $udd$ matter
is small.
By introducing heavy $s$ quarks,
the Fermi energy of $uds$ matter becomes lower than that of $ud$ matter.
Note that the Pauli potential gives a spurious potential energy to the system,
which should be renormalized into other effective potential terms \cite{Maru97}.
 One possibility to avoid this problem is, instead of using the Pauli potential,
 to maintain the Pauli principle of the system by stochastic rearrangements
of particle momenta \cite{Papa01}.
This model was originally developed for nucleon systems and was applied
also to the quark system very recently \cite{Terra04}.

\subsubsection{Meson exchange potentials}
\begin{figure}[b]
\resizebox{0.47\textwidth}{!}{%
\includegraphics{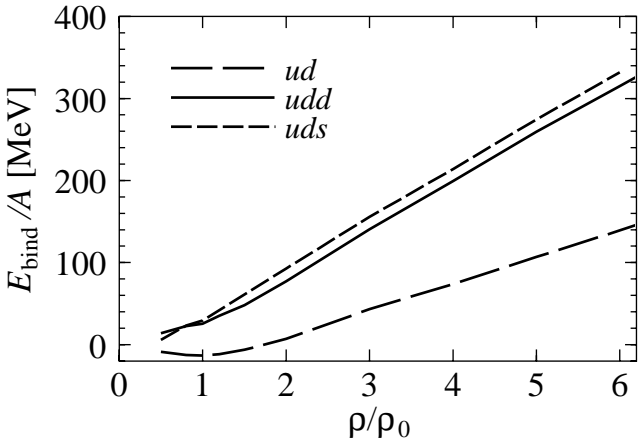}}
\caption{Binding energies per baryon for $ud$,
$udd$ and $uds$ matter
 calculated by the baryon cooling.
The saturation is seen for the $ud$ matter, but
it is not seen for the $udd$ and $uds$ matter.
}
\label{fig:2}       
\end{figure}
Though the meson exchange potentials do not affect
the baryon-quark transition, we adjust the meson exchange potentials
to discuss the EOS of strange and normal nuclear matters.
Prameters are determined to get appropriate ground state energy of ``baryon matter''
by frictional cooling with
a constraint that quarks form baryon clusters (baryonization constraint).
First we randomly distribute baryons which are composed of three quarks
and solve the cooling
equation with the baryonization constraint as,
\begin{eqnarray}
\dot{\bf R}_i&=&\frac{1}{3}\sum_{j\in \{i\}}\left[
\frac{\partial H}{\partial {\bf P}_j}
+\mu_r\frac{\partial H}{\partial {\bf R}_j} \right] ,
\nonumber \\
\dot{\bf P}_i&=&\frac{1}{3}\sum_{j \in \{i\}}\left[
-\frac{\partial H}{\partial {\bf R}_j}
+\mu_p\frac{\partial H}{\partial {\bf P}_j} \right] ,
\end{eqnarray}
here $\{i\}$ means a set of three quarks in a baryon
to which $i$-th quark belongs \cite{Maru00}.
We call this cooling ``baryon cooling''.
In Fig.\ 2 baryon density dependence of the energy per baryon is shown
for $ud$ matter (corresponding to symmetric nuclear matter),
$udd$ matter (neutron matter) and
$uds$ matter ($\Lambda$ matter).
Note that in this calculation with the baryonization constraint,
the color-dependent interactions
are exactly canceled between the white baryons.
For the $\sigma$ exchange potential, $g_{\sigma q}=3.09$,
$m_{\sigma}=400$ MeV, $L_{\sigma}=1.2$ fm and $\varepsilon=0.1$ are used and
$g_{\omega q}=4.98$, $m_{\omega}=782$ MeV and $L_{\omega}=0.7$ fm  for the $\omega$ ex
change and
$g_{\rho q}=9.0$, $m_{\rho}=770$ MeV and $L_{\rho}=1.2$ fm  for the $\rho$ exchange potential.
In order to reproduce the well-known properties of matter,
we have introduced the effective widths for each meson exchange term.
We fit
the binding energy of $ud$ matter
to 16.5 MeV/nucleon at $1\rho_0$.
The EOS of $\Lambda$ matter, which depends on the $\phi$ exchange potential
is unsettled yet \cite{Song03}.
In order to obtain the saturation of $uds$ matter,
we need to introduce an attractive $K$ exchange interaction
between $u$-$s$ or $d$-$s$ quarks.
However the $K$ meson exchange is prohibited according to the RMF model.
Here we have only a repulsive  $\phi$ exchange interaction
between $s$ quarks for simplicity.
The relevant parameters are $g_{\phi q}=\sqrt{2} g_{\omega q}=7.04$ \cite{Song03},
$m_{\phi}=1020$ MeV and $L_{\phi}=0.7$ fm.
Even if we use another value, e.g.\ $L_{\phi}\sim 1$ fm, the behavior of the
EOS is not changed noticeably.

\section{Results for the finite density system}\label{result}
Here we investigate the stability and the structure of quark matter in the ground stat
e
(zero temperature) for a wide range of density.
We solve the cooling
equations (\ref{cool}) with quark degrees of freedom.
We call this cooling ``quark cooling''
in contrast to the baryon cooling.
Snapshots of the ground state
of $uds$ matter with the width parameter (I) ($L$=0.43 fm) are displayed in Fig.\ 3.
Three quarks in red, green and blue
located within a distance $d_{\rm{cluster}}$
are considered to be confined as a baryon
\begin{eqnarray}
\left| {\bf R}_i-{\bf R}_j \right|&<&d_{\rm cluster}, \nonumber\\
\left| {\bf R}_j-{\bf R}_k \right|&<&d_{\rm cluster}, \nonumber\\
\left| {\bf R}_k-{\bf R}_i \right|&<&d_{\rm cluster}, \nonumber\\
\{i,j,k\} &=&\{{\rm red,green,blue}\}, \nonumber\\
\label{judge}
\end{eqnarray}
where we use $d_{\rm {cluster}}=0.7$ fm.
Quarks judged to be in a confined state
are plotted in white and
those deconfined in its own color.
At the normal baryon density $1\rho_0$ all quarks are confined as baryons.
As density increases, some quarks become deconfined
and the confined and the deconfined states
coexist around $2.5\rho_0$.
Most of the quarks are deconfined at high baryon density $\rho\geq 3\rho_0$.
\begin{figure}
\resizebox{0.47\textwidth}{!}{%
\includegraphics{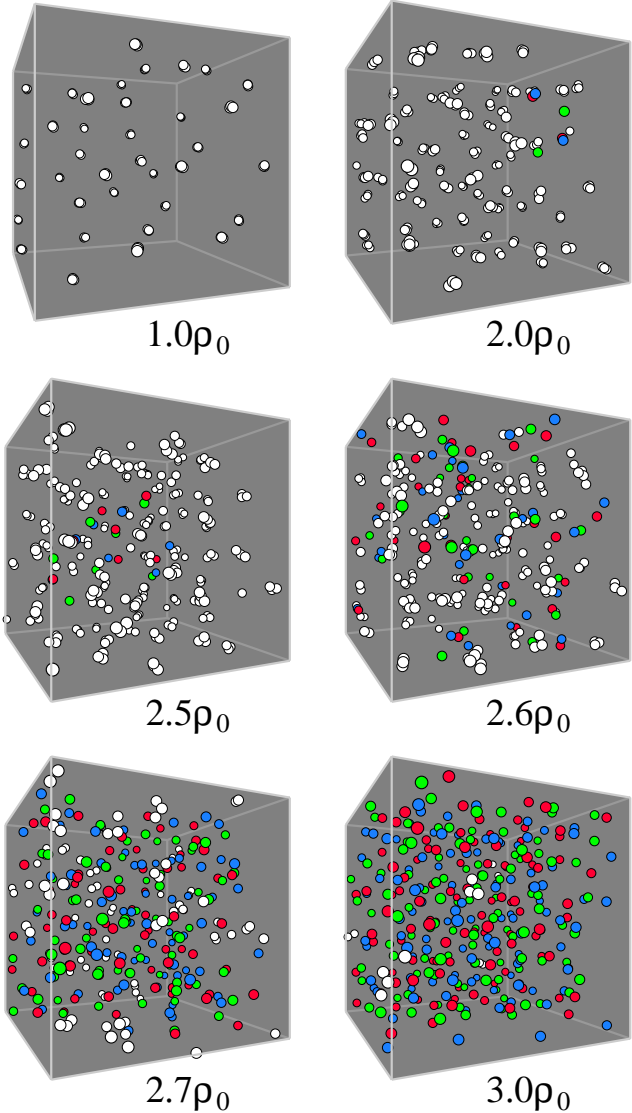}}
\caption{
Snapshots of $uds$ matter with $L$=0.43 fm at various densities.
Using the criterion of Eq.\ (\ref{judge}),
confined quarks are plotted with white and deconfined
quarks its own color.
At $1\rho_0$ all quarks are confined. As the density increases deconfined
quarks or its colors begin to appear. Most of the quarks are deconfined at $3\rho_0$.}
\label{fig:3}       
\end{figure}
\begin{figure}
\resizebox{0.47\textwidth}{!}{%
\includegraphics{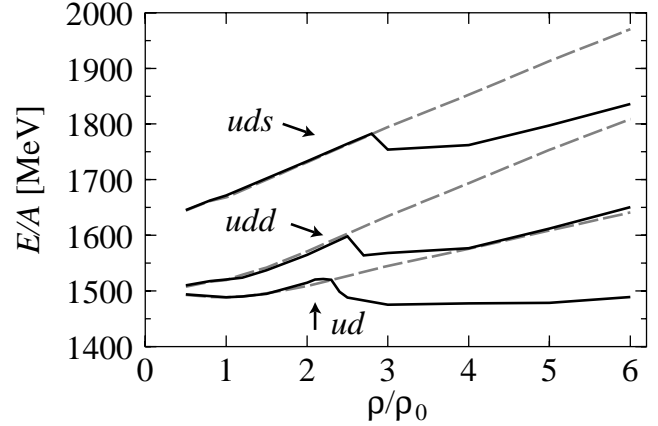}}
\caption{Density dependence of energy per baryon
for $ud$, $udd$ and $uds$ matter
in case of width parameter (I) ($L$=0.43 fm).
The dashed lines indicate the cases of baryon cooling and solid lines
quark cooling.
}
\label{fig:4}       
\end{figure}
\begin{figure}
\resizebox{0.47\textwidth}{!}{%
\includegraphics{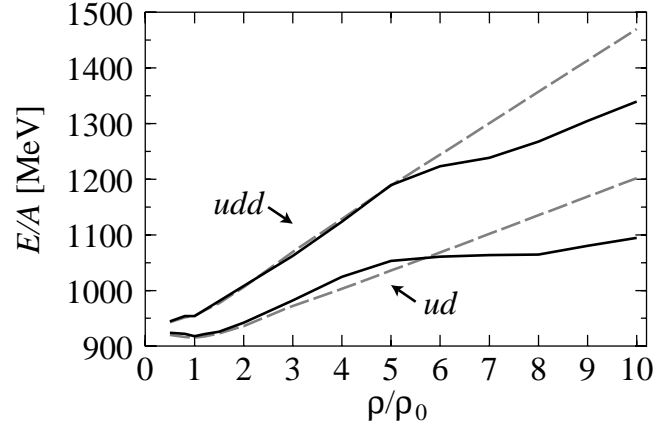}}
\caption{The same figure as in Fig.\ 4, but with
the width parameter (II) ($L$=0.33 fm).
}
\label{fig:5}       
\end{figure}
\begin{figure}
\resizebox{0.47\textwidth}{!}{%
\includegraphics{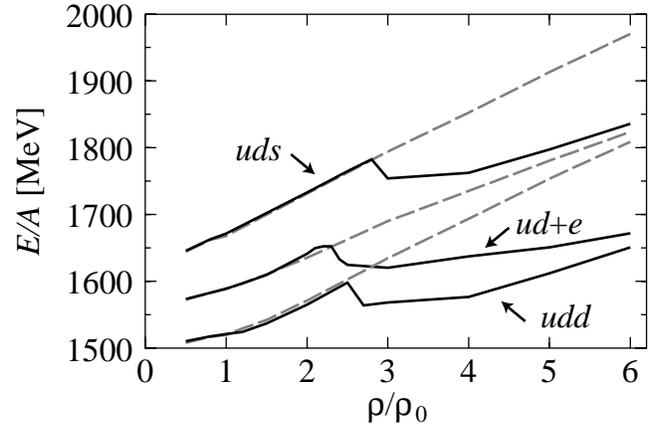}}
\caption{Same as Fig.\ 4, but under the electric charge-neutral condition.
}
\label{fig:6}       
\end{figure}
The ground state energies of $ud$, $udd$ and $uds$ matters with
the width parameter (I)
are shown by solid lines in Fig.\ 4.
The vertical axis indicates the energy per baryon.
For comparison, the case of baryon cooling is shown by thin dashed lines.
The energy by quark cooling agrees with that by baryon cooling at low density.
This means that quarks are confined as baryons.
At a certain density,
a baryon-quark transition occurs and the energy decreases.
The mechanism of baryon-quark transition in our dynamical model
is as follows.
%
If confined quarks are released from a baryon,
their kinetic energy, most of which has been the zero point energy,
decreases.
On the other hand, the potential energy increases due to the confinement force.
At above a certain density, the increase of potential energy at deconfinement
gets smaller due to the existence of many other quarks in the environment.
In this way, the decrease of the kinetic energy
is superior to the increase of potential energy.

In our result with the width parameter (I),
the critical density is lower than that usually expected.
The ground state energies by the width parameter (II) ($L$=0.33 fm)
for $ud$ and $udd$ matters is shown in Fig.\ 5.
The critical point of the baryon-quark transition is higher than that given
by the width parameter (I).
This implies that the transition occurs due to the density overlap of
a quark with those in other baryons.

Figure 6 shows the energies for the three kinds of matters under
the electric-charge neutral condition with the width parameter (I).
The difference compared to Fig.\ 4 is that the energy of relativistic electron
is added to the energy of $ud$ matter.
Below $6\rho_0$ the energy of $udd$ matter is the lowest.
Judging from the behavior of the curves, however,
$ud$ matter becomes more stable than $udd$ matter above $6\rho_0$.
\begin{figure*}
\resizebox{\textwidth}{!}{%
\includegraphics{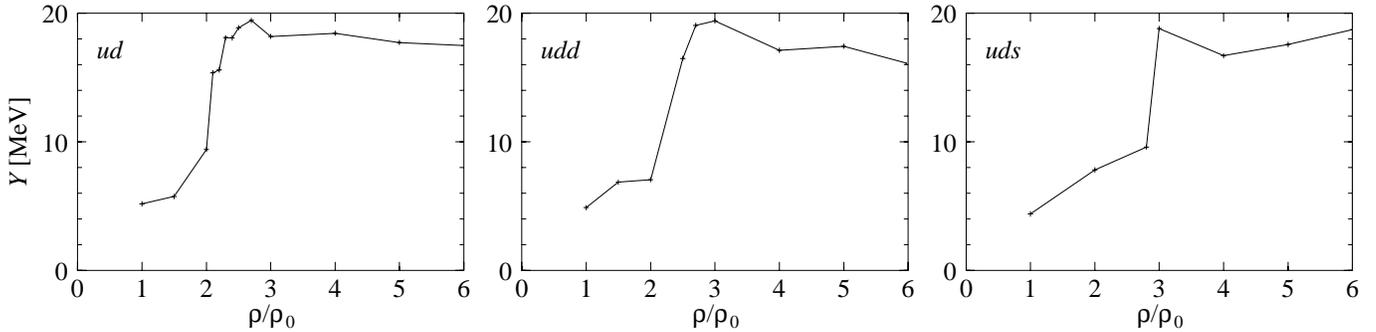}}
\caption{Density dependence of $Y$
for $ud$, $udd$ and $uds$ matter
in case of width parameter (I) ($L$=0.43 fm).}
\label{fig:7}       
\end{figure*}

The stability of strange matter has long been an attention of
many nuclear and astrophysicists \cite{Bodm71,witten}.
Our present result shows that $uds$ matter is
not favored even at high density.
However, we cannot give a definite statement
since our EOS of $uds$ matter has ambiguity caused by ignorance of the
$\Lambda$-$\Lambda$ interaction.
The width of $s$ quarks and interactions concerning $s$ quarks
influence the stability as well.

Here we have examined these three kinds of fixed flavor matter for simplicity.
For more realistic studies of matter realized in nature such as the core
region of neutron stars,
 however, a simulation of matter
in beta equilibrium is necessary.
%
%
The most stable matter with non-integer $u$-$d$-$s$ proportion
would be close to $udd$ matter at lower density,
and between $udd$ and $ud$ matter at higher density.
With the width parameter (I) there are density regions where
the slope of $E/A$ versus $\rho/\rho_0$ is negative in Fig.\ 6.
Generally such a situation is not realized by forming non-uniform structure.
In our case, however, the small size of the simulation box
and the introduction of density-dependent parameters (\ref{separation})
caused this problem.

Now let us define a quantity $Y$ which indicates the degree of deconfinement as,
particle feels.
\begin{equation}
Y=\left(\left<
\left|
\sum_i^{n}V_{\rm color}\left({\bf R}_i-{\bf R}\right)
\right|^2
\right>_{\bf R}\right)^{1/2}
\end{equation}
where {\bf R} is the position of a test particle in any color,
$V_{\rm color}$ is the color potential which the test particle feels
and $\left<\right>_{\bf R}$ means the average over ${\bf R}$
which is sampled for 10000 points.
If all quarks are confined in compact baryons,
the value of $Y$ becomes  zero.
Figure 7 shows the density dependences of $Y$.
The value of $Y$ is small at low density.
From $2\rho_0$ to 3$\rho_0$ which corresponds to
the density region of baryon-quark transition,
$Y$ increases gradually.
This gradual change of $Y$ does not necessarily indicate
the second order phase transition,
since the mixed phase in the first order phase transition may also
show the similar behavior of physical quantities.
\section{extension to the finite temperature}\label{tempera}
\begin{figure*}
\resizebox{\textwidth}{!}{%
\includegraphics{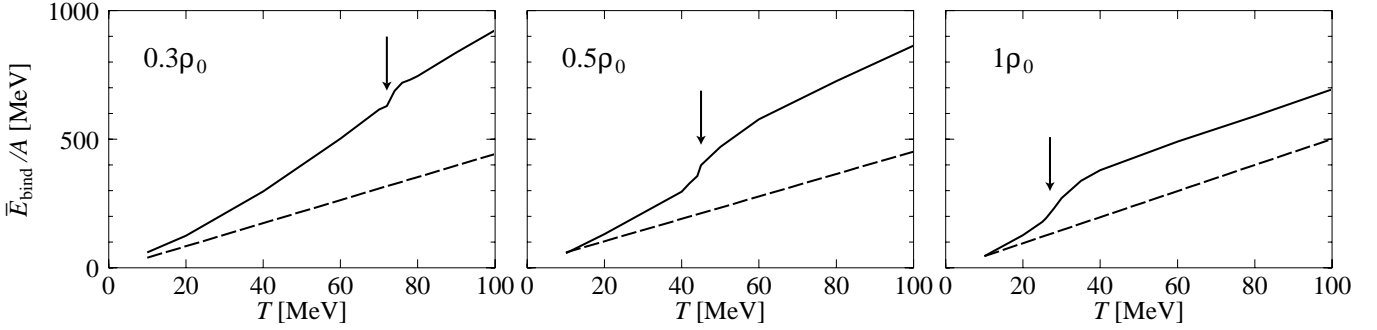}}
\caption{Averaged binding energy per nucleon for $ud$ matter with $L$=0.43 fm
at 0.3$\rho_0$ (left), 0.5$\rho_0$ (center) and 1$\rho_0$ (right)
are shown with solid line.
The dashed line denotes the case in which quarks are constrained to form baryons.
The arrows indicate the critical temperature.
}
\label{fig:8}       
\end{figure*}
\begin{figure*}
\resizebox{\textwidth}{!}{%
\includegraphics{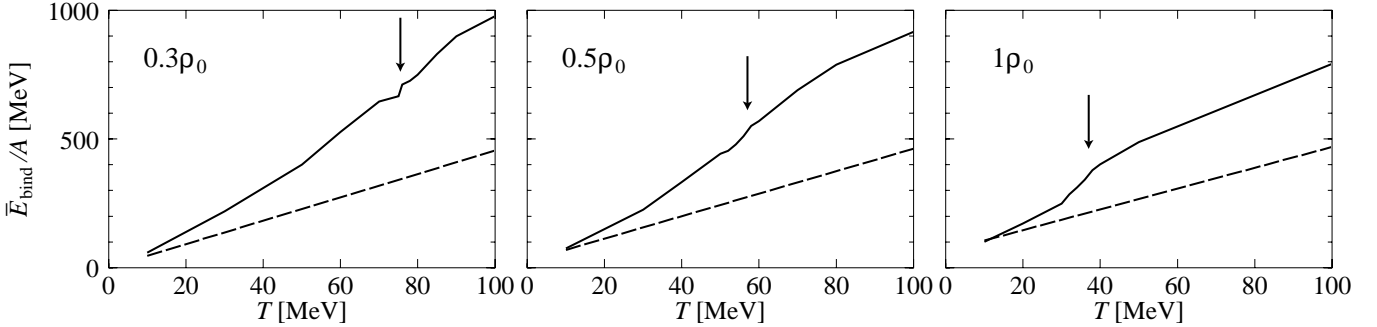}}
\caption{The same figure as in Fig.\ 8, but for $uds$ matter.
}
\label{fig:9}       
\end{figure*}
\subsection{Introduction of an effective temperature}
We extend above MD
to finite temperature systems.
First of all, the definition of a temperature commonly used
in MD is modified
from the common expression due to the momentum-dependent Pauli potential
according to Ref.\ \cite{Watnb},
\begin{equation}
\frac{3}{2}T_{\rm eff}=\frac{1}{n}\sum_{i=1}^{n}\frac{1}{2}{\bf P}_i \cdot
\left( \frac{\partial H}{\partial{\bf P}_i}\right).
\end{equation}
A popular method to control temperature in MD is
known as Nos\'e-Hoover method \cite{Hoo}.
The equilibrium is achieved by introducing an imaginary heat bath
which interacts with a real system
through an artificial coordinate and its momentum.
The extended Hamiltonian for thermostat systems is written as
\begin{equation}
H_{\rm Nose}=H+\frac{1}{2}Q{\xi}^2+3nT_{\rm set}\log{w},
\end{equation}
with $\xi=P_w/Q$, where $w$ is an artificial variable, $P_w$ is a
momentum conjugate to $w$ and $Q$ corresponds to its mass.
$T_{\rm set}$ is a target temperature of the system which is given as a parameter.
The equations of motion are
\begin{equation}
\dot{\bf R}_i=\frac{\partial H_{\rm Nose}}{\partial {\bf P}_i}
=\frac{\partial H}{\partial {\bf P}_i},
\end{equation}
\begin{equation}
\dot{\bf P}_i=-\frac{\partial H_{\rm Nose}}{\partial {\bf R}_i}
=-\frac{\partial H}{\partial {\bf R}_i}-\xi{\bf P}_i,
\end{equation}
\begin{equation}
\dot{w}=\frac{\partial H_{\rm Nose}}{\partial P_w}
=\xi,
\end{equation}
\begin{equation}
\dot{\xi}=\frac{3n}{Q}\left(T_{\rm eff}-T_{\rm set} \right).
\end{equation}
Here $H_{\rm Nose}$ is conserved
but the original effective Hamiltonian $H$ fluctuates.
The value of $T_{\rm eff}$ also fluctuates around $T_{\rm set}$ by the thermostat
friction  $\xi$.
\subsection{Results for the finite temperature system}\label{retem}
We investigate the thermal property of quark matter
for a wide range of density and temperature.
Figure 8  shows temperature dependence of the averaged energy
per baryon for $ud$ matter
at $0.3\rho_0$, $0.5\rho_0$ and $1\rho_0$ with $L=0.43$ fm from the left
by solid lines.
Simulation results with the baryonization constraint are also shown by dashed lines
for comparison.
A baryon-quark transition can be seen as the change of specific heat.
We define the critical temperatures by the maximum-specific-heat points
and are indicated by arrows in this figure.
The critical temperature becomes small as density increases.
The lines without the baryonization constraint (normal calculation) disagree with
those with the baryonization constraint
because quarks can move in the baryon in case of normal calculations.
The same quantity but for $uds$ matter is shown in Fig.\ 9.
The critical temperatures for $uds$ matter are 5\ --\ 15 MeV larger
than that for $ud$ matter at the same density.
This is due to the presence of the $s$ quarks having a larger mass
than $u$ and $d$ quarks.

We draw a phase diagram by using our results
of $ud$ matter for $L$=0.43 fm in Fig.\ 10 and
for $L$=0.33 fm in Fig.\ 11.
Baryon phase is realized at lower side of density and temperature, and quark phase is
realized
at higher side
in both diagrams, but baryon phase in Fig.\ 11 is larger than that of Fig.\ 10.
This means that the quark width affects
the phase transition due to the increase of temperature as well as the increase of den
sity.

\section{Summary}\label{summ}

Quark many-body systems were studied by MD where
the ground state was defined in a definite manner in terms of a Pauli potential.
The EOS were reproduced for three kinds of baryon matters with effective meson exchange
interactions between quarks.
Baryon-quark transition is seen when baryon density increases.
We have used two quark widths, $L=0.43$ fm and $L=0.33$ fm.
The density at which baryon-quark phase transition occurs is different
for the two widths since the transition is caused by the overlap of the quarks.
In case of the larger value of $L=0.43$ fm
the baryon-quark transition occurs at rather low density, around $3\rho_0$.
For $L=0.33$ fm case, the transition occurs around $5\rho_0$,
which is consistent with other theoretical calculations.
\begin{figure}[t]
\resizebox{0.47\textwidth}{!}{%
\includegraphics{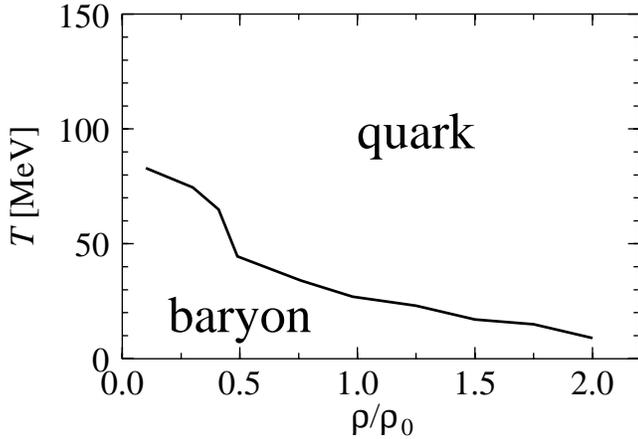}}
\caption{Phase diagram with the density and temperature for $ud$ matter
in case of $L$=0.43 fm.
}
\label{fig:10}       
\end{figure}
\begin{figure}
\resizebox{0.47\textwidth}{!}{%
\includegraphics{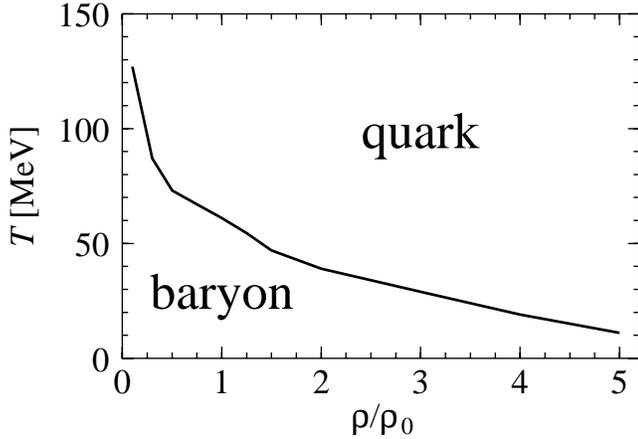}}
\caption{Phase diagram with the density and temperature for $ud$ matter
in case of $L$=0.33 fm.
}
\label{fig:11}       
\end{figure}

We have compared three kinds of matters with different
$u$-$d$-$s$ compositions: $ud$, $udd$ and $uds$.
Our results show that $ud$ matter is the most stable among them
and strange $uds$ matter is not stable.
For the EOS of $uds$ or $\Lambda$ matter, there would be still
room for improvement to take into account the future experiment.
For the color interaction,
the color magnetic interaction is necessary as a color and flavor dependent
interaction.
The medium effects of constituent quark masses by the chiral symmetry
restoration are also important to discuss the stability.
A possibility to include the medium effect is to use
the density dependent quark mass model (DDQM) \cite{DDQM}. However, the quark mass is
derived there from the global density and is common for all quarks.
This prescription may be too simple
since the individual particle motion is essential in MD.
A similar discussion can be done for the coupling constant $\alpha_s$.

We have extended our model for the finite temperature systems.
The baryon-quark transition can be seen as a change of specific heat.
The critical temperature for $uds$ matter is higher than $ud$ matter
by 5 -- 15 MeV due to the heavy mass of $s$ quark.
Like the critical density,
the critical temperature gets higher in case of $L$=0.33 fm
than that of $L$=0.43 fm.

It is necessary to include $\bar{q}q$ creation/annihilation process
for high temperature.
Lack of antisymmetrization and the asymptotic freedom are also open problems.

\begin{acknowledgement}

Y.~A.\ is grateful to T.~Tanigawa, H.~Koura, T.~Tatsumi,
T.~Hatsuda, Y.~Maezawa and T.~Endo
for their valuable discussions.

\end{acknowledgement}

\end{document}